\documentclass{webofc}

\usepackage[varg]{txfonts}   
\usepackage{hyperref}
\usepackage{url}
\newcommand{\heff}{$H_{\rm eff}$}
\newcommand{\heffs}{$H_{\rm eff}$s}
\newcommand{\qbox}{$\hat{Q}$~box}
\hypersetup{colorlinks=true,citecolor=blue,urlcolor=blue,linkcolor=blue}
\begin{document}
\title{Nuclear structure study with two- and three-nucleon contact
  interactions derived within low-energy EFT}
%
%

\author{\firstname{Songlin} \lastname{Lyu}\inst{1,2}\fnsep\thanks{\email{songlin.lyu@unicampania.it}} \and
  \firstname{Francesco} \lastname{Amodio}\inst{1} \and
  \firstname{Giovanni} \lastname{De Gregorio}\inst{1,2} \and
  \firstname{Nunzio} \lastname{Itaco}\inst{1,2} \and  
        \firstname{Luigi} \lastname{Coraggio}\inst{1,2}
}

\institute{Dipartimento di Matematica e Fisica, Universit\`a degli
  Studi della Campania ``Luigi Vanvitelli'', \\
  viale Abramo Lincoln 5 - I-81100 Caserta, Italy
\and
  Istituto Nazionale di Fisica Nucleare, Complesso Universitario
  di Monte  S. Angelo, \\ Via Cintia - I-80126 Napoli, Italy
}

\abstract{We present the results of the application of a nuclear
  potential consisting of two- and three-nucleon contact interactions
  in nuclear structure investigations.
  The nuclear Hamiltonian has been derived for a very low-energy
  regime within the framework of the effective field theory, its
  low-energy constants have been fitted to a few low-energy
  nucleon-nucleon experimental observables and the deuteron and $^3$H
  binding energies.
  Our goal is to validate the ability of this Hamiltonian to reproduce
  some important features of open-shell nuclei, and to this end we
  derive effective shell-model Hamiltonians for nuclei in the $p$- and
  $sd$-shell mass regions.
  The results of shell-model calculations with these effective
  Hamiltonians are then compared with experiment, and also with those
  obtained with a nuclear Hamiltonian derived within chiral
  perturbation theory, that includes also terms with one- and two-pion
  exchanges.
}
\maketitle
\section{Introduction}
\label{intro}
A major breakthrough in the last quarter of the century has been the
derivation of nuclear Hamiltonians consisting of two- and many-nucleon
components through procedures that are grounded on the application of
the effective field theory (EFT), as indicated in the seminal works of
S. Weinberg \cite{Weinberg90,Weinberg91}.

In particular, a great success has been achieved through the
construction of realistic two- and three-nucleon forces (2NF and
3NF) starting from a chiral Lagrangian, grounded on the concept of EFT
to study the $S$-matrix for processes involving any number of
low-momentum pions and nucleons.
In such a framework, the long-range forces are dictated by the
symmetries of low-energy QCD -- in particular the spontaneously broken
chiral symmetry -- and the short-range dynamics is absorbed by contact
terms that are proportional to low-energy constants (LECs), the latter
fitted on the observables related to the two- and three-nucleon systems.

A relevant feature of this approach to the derivation of the nuclear
Hamiltonian is that provides the framework for constructing nuclear
two- and many-body forces on an equal footing \cite{Machleidt11},
since most of the interaction vertices (and LECs) that appear in the
many-body forces also occur in the 2NF.

In an extreme approach to the problem of the nuclear interaction, one
may choose a very low-energy regime, where the pion mass represents
the heavy scale, and pions are integrated out so that the nuclear
Hamiltonian is made up only by contact terms between two or more
nucleons.
This is the so-called pionless EFT \cite{Bedaque02} and provides an
interesting framework since captures the rich structures in low-energy
regime of nuclear interaction, and has demonstrated remarkable
accuracy in reproducing the ground states of various nuclei with mass
numbers up to about $A=50$ \cite{Lu19}.

In this work, we analyze some features that characterize nuclear
structure properties of light- and medium-mass nuclei, as obtained
starting from a nuclear Hamiltonian recently derived within such a
low-energy EFT (LEEFT) \cite{Schiavilla21}.
In particular, we are interested to study the shell evolution in
nuclei belonging to $0p$- and $1s0d$-shell, a property that is linked
to the observed shell closures in nuclear structure.
To this end, we have derived effective shell-model (SM) Hamiltonians
(\heffs) by way of the many-body perturbation theory \cite{Coraggio20}
starting from the above mentioned LEEFT potential, accounting both for
its two- and three-body components, and comparing the calculated
closure properties with the observed ones, as well as with those
provided by another nuclear Hamiltonian derived through the chiral
perturbation theory (ChPT).
The latter is the well-known N$^3$LO 2NF potential derived by Entem
and Machleidt \cite{Machleidt11}, supplemented by a 3NF component
derived at N$^2$LO in ChPT consistently with the 2NF component.

This paper is organized as follows: in Sec. \ref{outline} we outline
the characteristics of the LEEFT potential we start from, as well as
the basics of the derivation of the effective SM Hamiltonian \heff.
Sec. \ref{results} is devoted to the presentation of the results we
have obtained with LEEFT \heff, and the comparison with experiment as
well as with those obtained with the Entem-Machleidt ChPT potential.
We focus our attention on the shell evolution in $0p$- and
$1s0d$-shell, and how it impacts on the observed closure properties.
In the last section (Sec. \ref{conclusions}), we summarize the
conclusions of this study and the assessment of the ability of LEEFT
to reproduce the observed shell evolution in open shell nuclei.

\section{Outline of the theory}\label{outline}
As mentioned in the Introduction, our study originates from the
nuclear Hamiltonian constructed within EFT as introduced in
Ref. \cite{Schiavilla21}, and that is characterized by both 2NF and
3NF components, built up only through contact terms in a perturbative
expansion of the EFT Lagrangian.

In Ref. \cite{Schiavilla21} different combinations of Gaussian cutoffs
are considered in the construction of LEEFT Hamiltonians at different
orders of the perturbative expansion of the EFT Lagrangians, and we
have chosen for our calculations the one that is labelled as {\it
  ``optimized''}.
For the 2NF component of the nuclear Hamiltonian, we consider both the
one derived at the leading order (LO) of the perturbative expansion,
as well as the one at next-to-next-to-next-to-leading order (N$^3$LO).
As regards the 3NF component, only the one at LO has been derived,
which includes just a single contact term (see Ref. \cite{Schiavilla21}).

Starting from this nuclear Hamiltonian we have constructed effective
SM Hamiltonians for nuclei belonging to the mass region of $0p$ and
$1s0d$ shells, and to this end we derive \heff~ by way of the
time-dependent perturbation theory \cite{Kuo90}.
Namely, \heff~ is expressed through the Kuo-Lee-Ratcliff (KLR)
folded-diagram expansion in terms of the vertex function \qbox, which
is composed of irreducible valence-linked diagrams \cite{Kuo71}.

We include in the perturbative expansion of the \qbox~ one- and
two-body Goldstone diagrams up to third order for those accounting for
the 2NF vertices, and up to first order for the one with a 3NF vertex.
A detailed description of the calculation of the diagrammatic series
for the derivation of \heff, including 2NF and 3NF vertices, is
presented in Refs. \cite{Coraggio20,Coraggio12}.

This is the so-called realistic shell-model approach (RSM) to the
derivation of SM effective Hamiltonians \cite{Coraggio20}

It is worth noting that, in a SM approach, the effective Hamiltonian
for one valence-nucleon systems provides the SP energies for the SM
calculations, then the two-body matrix elements (TBME) of the SM
Hamiltonian are obtained by way of a subtraction procedure from \heff~
for nuclei with two valence nucleons outside the doubly-closed core.

Actually, SM studies involve nuclear systems are characterized by a
number of valence nucleons that is larger than one and two, then
effective many-body forces --  arising from the interaction via the
two-body force of the valence nucleons with excitations outside the
model space -- have to be included in the perturbative expansion of
\heff, besides the 3NF component of the nuclear Hamiltonian.

To meet this issue, we also account for the contributions of three-body
diagrams calculated at second order in perturbation theory, and, since
the code we employ diagonalize the SM Hamiltonian cannot manage
three-body components of the residual interactions, we derive
density-dependent two-body contribution at one-loop order from these
three-body diagrams, summing over the partially-filled model-space
orbitals \cite{Coraggio20}.

\section{Results}\label{results}
\begin{figure}[h]
\centering
\includegraphics[width=8.5cm,clip]{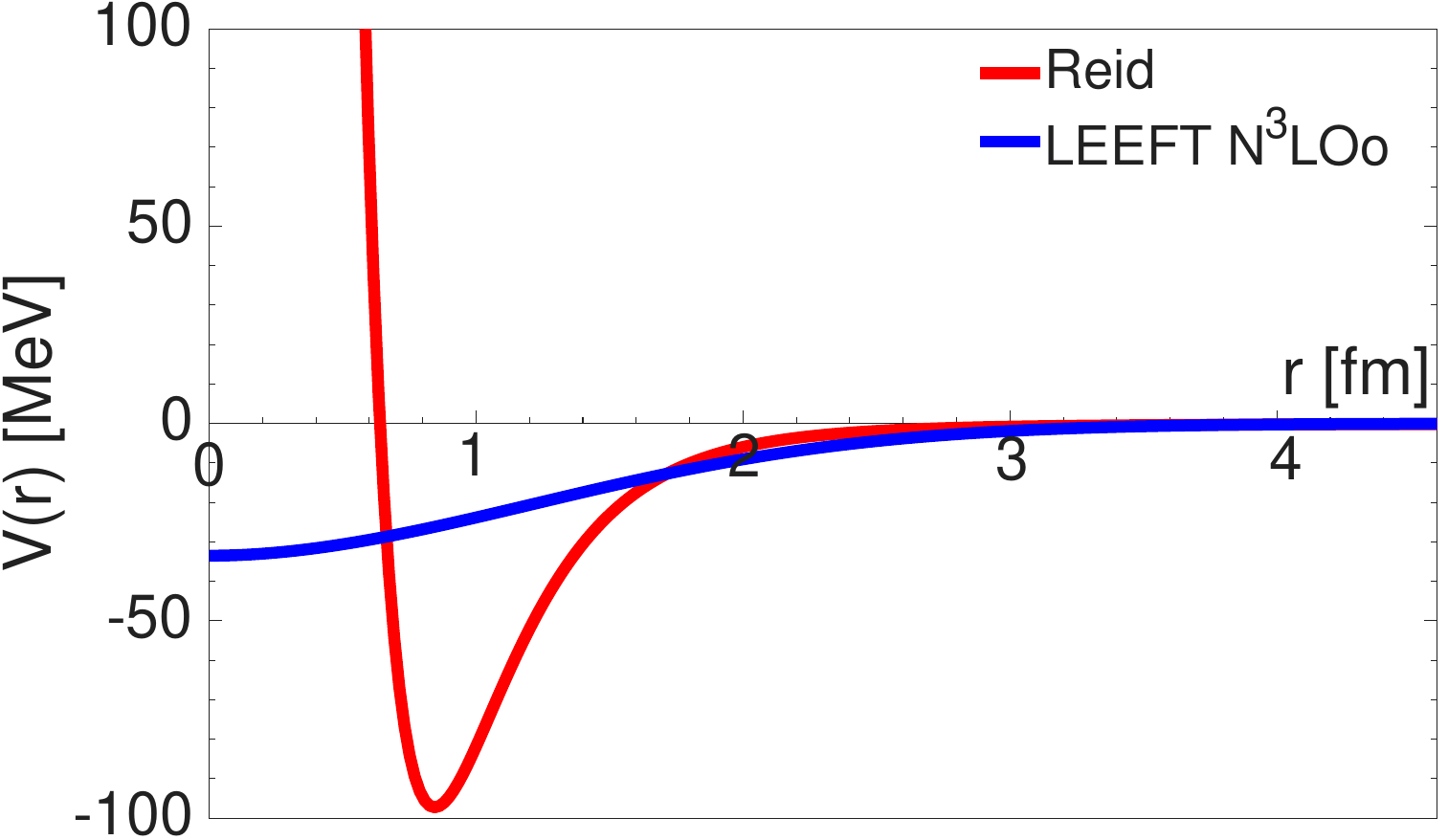}
\caption{Behavior of 2NF for LEEFT potential at N$^3$LO and Reid
  interaction in the $^1{\rm S}_0$ channel.}
\label{NN1S0}
\end{figure}

As a starting point of our analysis, it is worth comparing the 2NF
matrix elements of the LEEFT Hamiltonian with those of a realistic
nucleon-nucleon ($NN$) potential that is constructed to reproduce at
the best the data of the $NN$ scattering, as well as the
deuteron properties.
To this end, we compare in Fig. \ref{NN1S0} the behavior of 2NF
component of LEEFT Hamiltonian at N$^3$LO (optimized parametrization)
in the $1{\rm S}_0$ channel with the corresponding one of the Reid
potential \cite{Reid68}, as a function of the $NN$ relative distance.

It is worth pointing out a specific detail that should be considered
before discussing the extraordinary difference between the behavior of
the 2NF for these nuclear Hamiltonians.
The Reid potential has been constructed to reproduce the $NN$
scattering phase-shifts up to 350 MeV laboratory energy, then it is
sensitive to the strong repulsion at short $NN$ distances, which shows
up as a change of sign of the $NN$ phase shifts in the $s$-wave
channels around 250 MeV laboratory energy.
On the other side, the LEEFT Hamiltonian is constructed within EFT by
considering the pion mass as the heavy scale, then its 2NF component
cannot be sensitive to the observed short-range repulsion.
As a matter of fact, the LO 2NF potential reproduces singlet $np$
scattering length and deuteron binding energy,
the NLO potential reproduces $NN$ scattering data up to 15 MeV
laboratory energy, and the N$^3$LO 2NF one up to 25 MeV
\cite{Schiavilla21}.

These characteristics explain the different short-range behavior 
of the two 2NF potentials -- Reid potential is strongly repulsive,
whereas LEEFT is attractive --, and, as we will see, is partly
responsible for some features of the LEEFT Hamiltonian in the
description of the shell evolution of medium-mass nuclei.

\begin{figure}[h]
\centering
\includegraphics[width=10cm,clip]{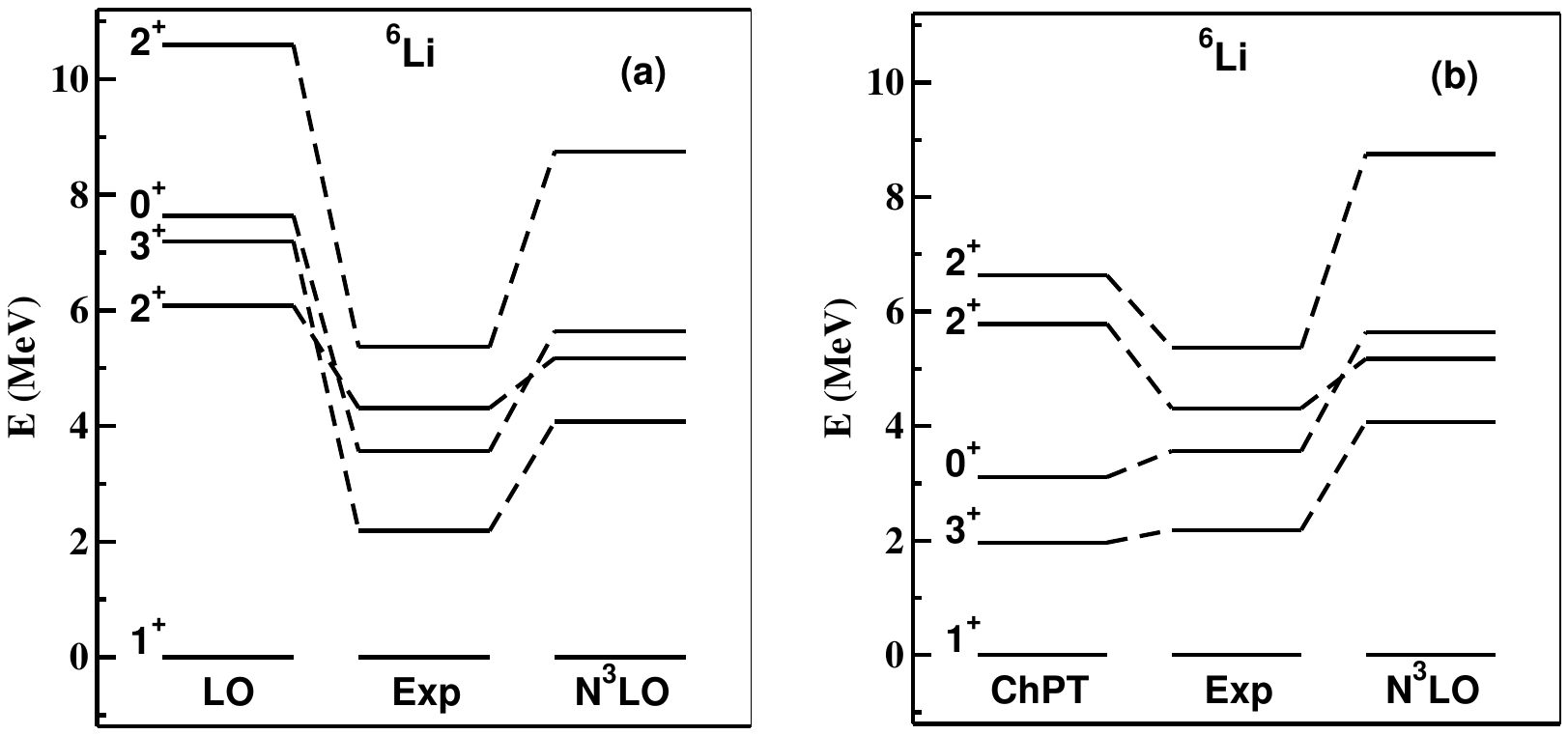}
\caption{In panel (a), calculated spectra of $^{6}$Li with LO and
  N$^3$LO LEEFT effective SM Hamiltonian compared with experiment. In
  panel (b), the same as in (a) but for ChPT and N$^3$LO \heffs.}
\label{6Li}
\end{figure}

In panel (a) of Fig. \ref{6Li} they are reported the excitation spectra of
$^6$Li, calculated starting from LEEFT Hamiltonians with 2NFs at LO
and at N$^3$LO and a LO 3NF component for both cases, as mentioned in
Sec. \ref{outline}, by way of \heffs~ defined in the $p$ shell and
constructed following the procedure reported in Sec. \ref{outline}.
The theoretical spectra are compared with the experimental one
\cite{ensdf}, and we may observe that both LEEFT \heffs~ provide a
poor reproduction of the low-lying excited states, with a few
inversions in the energy-level sequence.
This feature is even clearer if we compare in panel (b) the results
obtained with LEEFT \heff~ with those obtained by deriving \heff~ from
ChPT Hamiltonian, namely a N$^3$LO 2NF and a N$^2$LO 3NF
\cite{Machleidt11}, as in Ref. \cite{Fukui18}.
As can be seen, the ChPT \heff~ provides a nice reproduction of the
experimental spectrum.

\begin{figure}[h]
\centering
\includegraphics[width=10cm,clip]{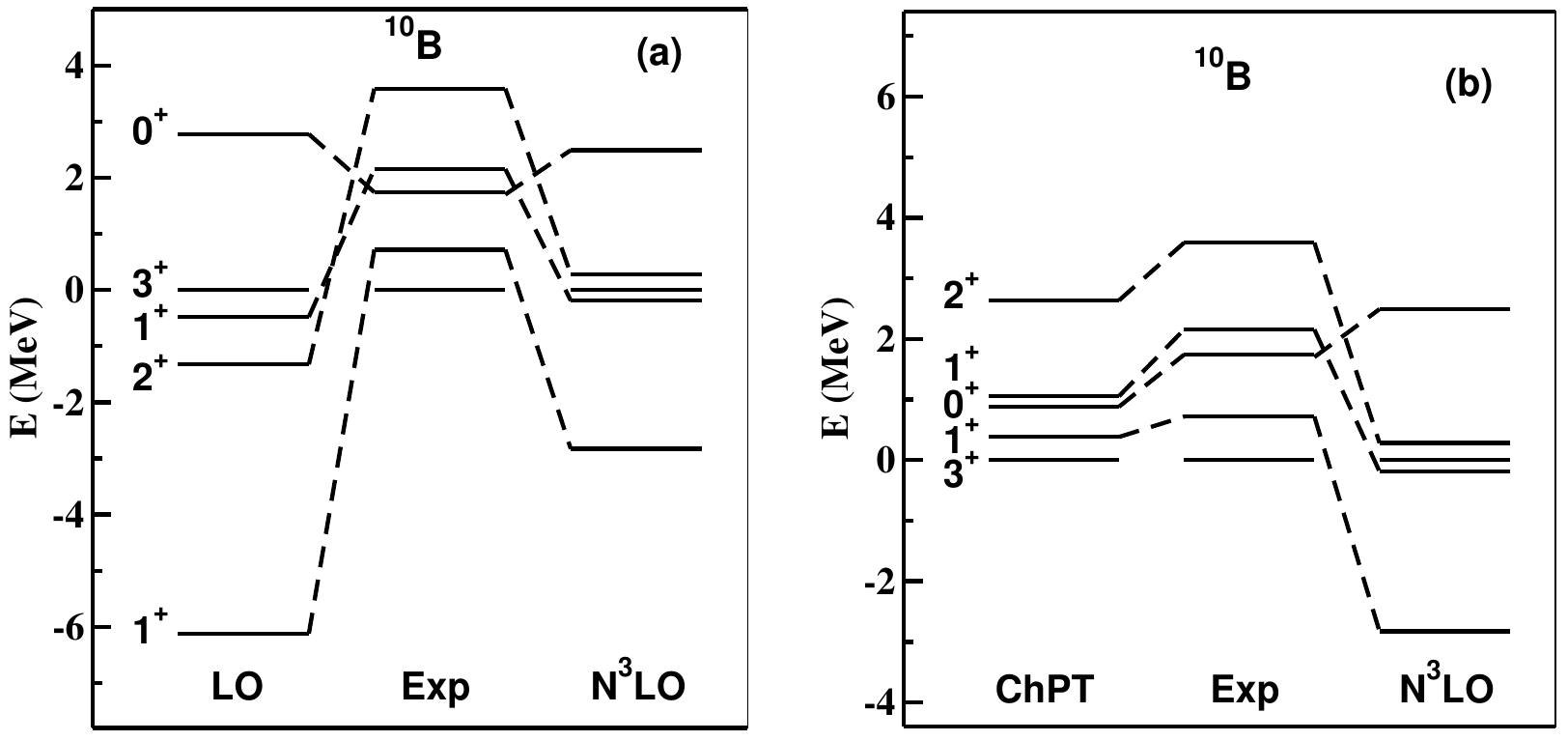}
\caption{Same as in Fig. \ref{6Li}, but for $^{10}$B}
\label{10B}
\end{figure}

The failure of the results obtained with LEEFT \heff~ is even more
evident if we consider the low-lying excited states of $^{10}$B and as
they are reproduced by LO and N$^3$LO \heffs, as reported in
Fig. \ref{10B}.
The calculated ground state is the yrast $J^{\pi}=1^+$ one, that is
strongly depressed with respect to the yrast $J^{\pi}=3^+$ which
experimentally is the ground state.
There is a strong contrast with the one calculated with ChPT \heff,
reported in panel (b), that reproduces correctly both the experimental
sequence of the excited states as well as the observed energy density.
It is worth pointing out that Navratil {\it et al} in their study of
$p$ shell nuclei \cite{Navratil07} pinned the role of the 3NF
component of the nuclear Hamiltonian in reproducing correctly the
ground state of $^{10}$B.
This may lead to the hypothesis that the 3NF component of the LEEFT
Hamiltonian, that currently consists only of a contact term at LO, is
not sufficiently adequate to balance the strong attractive behavior of
the 2NF component and describe satisfactorily the low-lying spectra in
$p$ shell nuclei.

We conclude the analysis of the results for $p$-shell region by
considering the shell evolution as reproduced by LEEFT \heffs.

Shell evolution and subshell closures are mainly driven by the energy
spacings of SP states.
The SP energies of $0p_{3/2},0p_{1/2}$ orbitals with LO \heff~ are
degenerate, while with N$^3$LO \heff the SP spacing is about 3 MeV for
both proton and neutron orbitals.
Then one may expect that LO \heff~ cannot be able to reproduce shell
closures appearing with the filling of the $0p_{3/2}$ orbitals, while
the N$^3$LO might be able to provide a better comparison with
experiment.

\begin{figure}[h]
\centering
\includegraphics[width=10cm,clip]{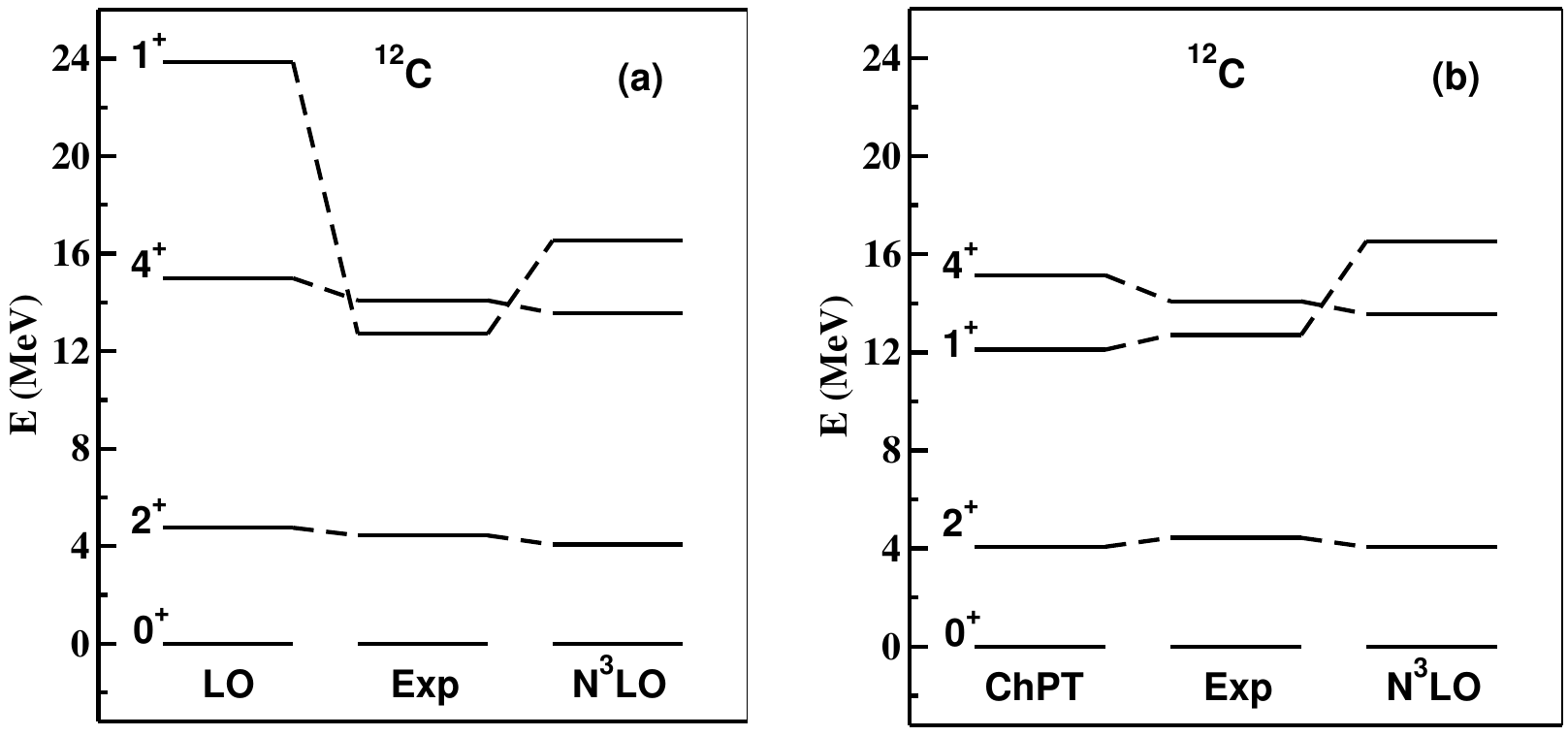}
\caption{Same as in Fig. \ref{6Li}, but for $^{12}$C}
\label{12C}
\end{figure}

In order to verify the shell evolution properties of these two \heffs,
in Fig. \ref{12C} we report the calculated and experimental low-lying
spectra of $^{12}$C, a nucleus that should correspond, in a
non-interacting SM picture, to a complete filling of the proton and
neutron $0p_{3/2}$ orbitals.

As can be seen in panel (a), both LEEFT \heffs~ reproduce nicely the
yrast $J^{\pi}=2^+,4^+$ states, but a more or less large inversion for
the yrast $J^{\pi}=1^+$ state.
This leads to the conclusion that the monopole component of the LO
\heff~ is able to enlarge the splitting of the effective SP energies
as a function of the number of valence nucleons (see the discussion in
Ref. \cite{Fukui18}), and provides the observed shell closure connected
with the energy spacing of the yrast $J^{\pi}=2^+$ state.

We now consider nuclei in the $1s0d$-shell mass region, and we spot
our attention on the oxygen isotopes.

\begin{figure}[h]
\centering
\includegraphics[width=10cm,clip]{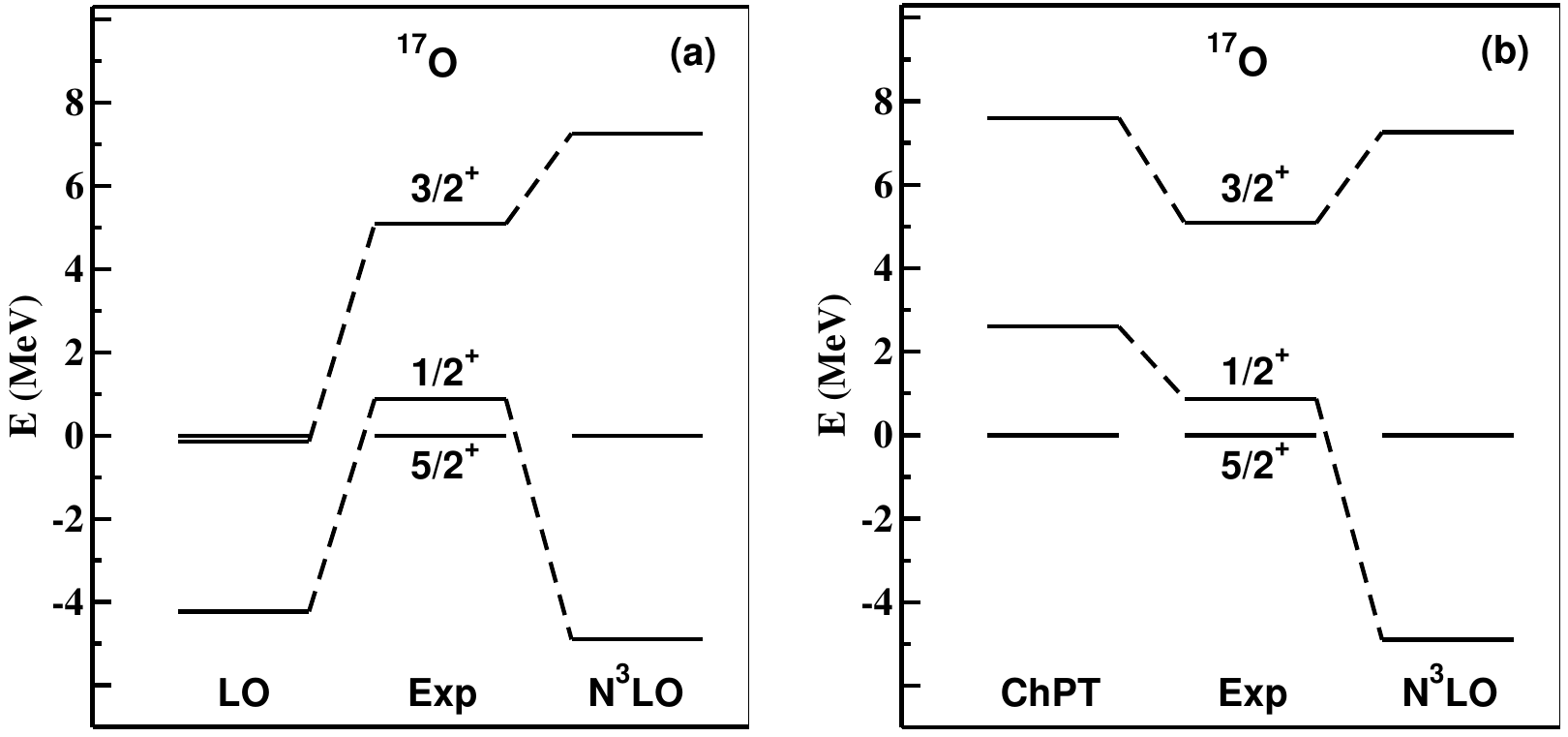}
\caption{In panel (a), calculated SP spectra of $^{17}$O with LO and
  N$^3$LO LEEFT effective SM Hamiltonian compared with experimental
  states with the largest SP component. In panel (b), the same as in
  (a) but for ChPT and N$^3$LO \heffs.}
\label{17O}
\end{figure}

As before, we are interested in the shell evolution of this well-known
isotopic chain, then it is worth to start from the structure of the SP
spectrum of $^{17}$O, that is reported in Fig. \ref{17O}.
As shown in panel (a), there is a first unpleasant surprise that is
the wrong sequence of the SP levels obtained with the LEEFT \heffs,
which predict the neutron $1s_{1/2}$ as ground-state orbital.
We note that the latter is very depressed with respect to the $0d$
spin-orbit partners, at variance with respect to the observed SP
states \cite{ensdf}.
In particular, the LO \heff~ does not provide the observed
$0d_{5/2},0d_{3/2}$ spin-orbit splitting, while the N$^3$LO \heff~ is
at least able to reproduce this fundamental feature.
On the other side, the results with the ChPT \heff~ \cite{Lyu25}
reproduce nicely the experimental SP sequence.

\begin{figure}[h]
\centering
\includegraphics[width=9cm,clip]{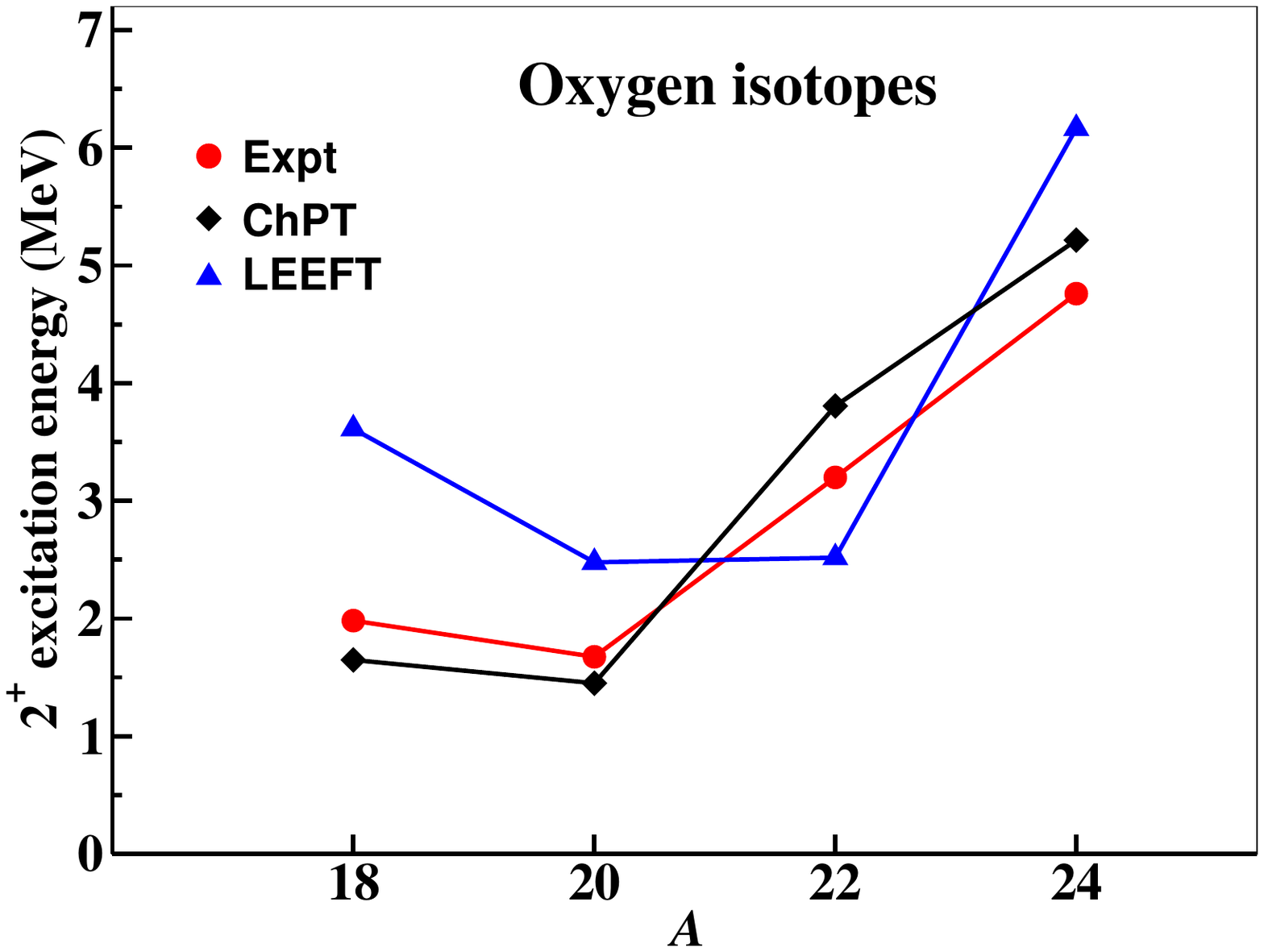}
\caption{Experimental and calculated excitation energies of the yrast
  $J^{\pi}=2^+$ states for even-mass oxygen isotopes from $A = 18$ to
  24 with ChPT and LEEFT \heffs.}
\label{J2p}
\end{figure}

The calculated SP spectrum, that represents the SP energies of the
shell model Hamiltonian, influence strongly the shell evolution of the
oxygen isotopes, as testified in Fig. \ref{J2p} by the excitation
energies of the yrast $J^{\pi}=2^+$ states for even-mass oxygen
isotopes as a function of the mass number $A$.

The observed raise of the excitation energy of the yrast $J^{\pi}=2^+$
state between $A=20$ and 22 corresponds to the filling of the neutron
ground-state $0d_{5/2}$ orbital, that is correctly reproduced by the
ChPT orbital but not by the LEEFT \heff.
This is related to a large collectivity of the $J^{\pi}=2^+_1$ wave
function, because LEEFT \heffs~ provide the $1s_{1/2}$ as ground state
in the SP sequence and a larger configuration mixing.
The unrealistic position of the $1s_{1/2}$ orbital is also responsible
of the artificial subshell closure in $^{18}$O, whose $J^{\pi}=2^+$
excitation energy is almost twice the observed one.

\section{Summary and conclusions}\label{conclusions}
In this contribution, we have presented the results of shell-model
calculations, where the SM effective Hamiltonian has been derived
through the many-body perturbation theory \cite{Coraggio20}, starting
from a nuclear Hamiltonian with two- and three-body components that
has been constructed by way of pionless EFT \cite{Schiavilla21}.

In particular, we have focused our attention on the issue of the shell
evolution, then comparing the calculated low-lying spectra of nuclei
in $0p$- and $1s0d$-shell mass regions with the experimental ones, in
order to evidence specific features that help to grade the ability of
these nuclear Hamiltonian in reproducing relevant nuclear-structure
observables.

As a matter of fact, our calculations unveil the poor agreement
with data for most of the nuclear systems we have considered, and, in
particular, provide several inversions with respect to the observed
sequence of the excited states.
It is worth mentioning the case of the oxygen isotopes, where the
incorrect sequence of the calculated SP energies with respect to the
experimental SP states of $^{17}$O ignite a distorted description of
the shell evolution of the oxygen isotopic chain.

These results may be mainly addressed to the fact that the nuclear
Hamiltonian under study has been derived within a very low-energy
scale, the pion mass representing the heavy scale.
In a previous investigation \cite{Coraggio15}, it has been evidenced
that, for nuclear Hamiltonians which are constructed within an
increasing low-energy regime, the role of many-body components, aside
the 2NF one, becomes more and more important in order to reproduce the
observed shell evolution.

This leads to our conclusion that in an extreme low-energy EFT, such
as the pionless EFT, in order to obtain a satisfactory agreement with
experimental shell evolution, the weight of many-body forces (3NF,
4NF, ...) in the construction of the nuclear Hamiltonian increases
with the mass of nuclear system under consideration.

\section{Acknowledgments}
This work was supported in part by the Italian Ministero
dell’Universit\`a e della Ricerca (MUR), through the grant Progetto di
Ricerca d'Interesse Nazionale 2022 (PRIN 2022) ``Exploiting Separation
of Scales in Nuclear Structure and Dynamics'', CUP B53D23005070006.


\begin{thebibliography}{}
\bibitem{Weinberg90} S. Weinberg, Phys. Lett. B {\bf 251}, 288 (1990).
\bibitem{Weinberg91} S. Weinberg, Nucl. Phys. B {\bf 363}, 3 (1991).
\bibitem{Machleidt11} R. Machleidt and D. R. Entem, Phys. Rep. {\bf 503}, 1
  (2011).
\bibitem{Bedaque02} Paulo F. Bedaque and Ubirajara van Kolck,
  Annu. Rev. Nucl. Part. Sci. {\bf 52}, 339 (2002).
\bibitem{Lu19} Bing-Nan Lu, Ning Li, Serdar Elhatisari, Dean Lee,
  Evgeny Epelbaum, and Ulf-G. Meißner, Phys. Lett. {\bf 797}, 134863 (2019).
\bibitem{Schiavilla21} R. Schiavilla, L. Girlanda, A. Gnech,
  A. Kievsky, A. Lovato, L. E. Marcucci, M. Piarulli, and M. Viviani,
  Phys. Rev. C {\bf 103}, 054003 (2021).
\bibitem{Coraggio20} Luigi Coraggio and Nunzio Itaco, Frontiers in
  Physics {\bf 8}, 345 (2020).
\bibitem{Kuo90} T. T. S. Kuo and E. Osnes, Lecture Notes in Physics
  {\bf 364}, Springer-Verlag Berlin (1990).
\bibitem{Kuo71} T. T. S. Kuo, S. Y. Lee, and K. F. Ratcliff,
  Nucl. Phys. A {\bf 176}, 65 (1971).
\bibitem{Coraggio12} L. Coraggio, A. Covello, A. Gargano, N. Itaco,
  and T. T. S. Kuo, Ann. Phys. (NY) {\bf 327}, 2125 (2012).
\bibitem{KSHELL} Noritaka Shimizu, Takahiro Mizusaki, Yutaka
  Utsuno, and Yusuke Tsunoda, Computer Physics Communications {\bf
    244}, 372 (2019).
\bibitem{Reid68} R. V. Reid, Ann. Phys. (N.Y.) {\bf 50}, 411 (1968).
\bibitem{ensdf} Data extracted using the NNDC On-line Data Service
  from the ENSDF database, https://www.nndc.bnl.gov/ensdf.
\bibitem{Fukui18} T. Fukui, L. De Angelis, Y. Z. Ma, L. Coraggio,
  A. Gargano, N. Itaco, and F. R. Xu, Phys. Rev. C {\bf 98}, 044305 (2018).
\bibitem{Navratil07} P. Navr\'atil, V. G. Gueorguiev, J. P. Vary,
  W. E. Ormand, and A. Nogga, Phys. Rev. Lett. {\bf 99}, 042501 (2007).
\bibitem{Lyu25} S. L. Lyu, G. De Gregorio, T. Fukui, N. Itaco, and
  L. Coraggio, Phys. Rev. C {\bf 112}, 054314 (2025).
\bibitem{Coraggio15} L. Coraggio, A. Gargano, and N. Itaco, JPS
  Conf. Proc. {\bf 6}, 020046 (2015).
  
\end{thebibliography}
\end{document}